\documentclass[12pt]{article}
\input{epsf.tex}

\begin{document}

\begin{center}
{\bf ESTIMATION OF MAXIMUM PERMISSIBLE ERRORS IN THE TOTAL GAMMA-SPECTRA INTENSITIES
AT DETERMINATION FROM THEM OF LEVEL DENSITY AND RADIATIVE STRENGTH FUNCTIONS} \end{center}
\begin{center}
{\bf  A.M. Sukhovoj, V.A. Khitrov}\\
\end{center}\begin{center}
{\it  Joint Institute
for Nuclear Research, 141980, Dubna, Russia}\\
\end{center}
\begin{abstract}
From a comparison of the total gamma-spectra calculated for different functional dependencies of level density and radiative strength functions, there were obtained both their square root relative differences and analogous data for the used parameters. The analysis of these data showed that the total uncertainty in determination of gamma-spectra intensities which is necessary to obtain reliable values of parameters of cascade gamma-decay, most probably, must not exceed one percent.

\end{abstract}

\section{Introduction}\hspace*{16pt}
The authors of the last variant  \cite{NIM} of known \cite{Bar} method for determination of level density  $\rho$ and radiative strength functions 
$k=f/A^{2/3}=\Gamma/(E_\gamma^3 D_\lambda A^{2/3})$ from total gamma-spectra of reactions  $^3$He,$(\alpha\gamma)$ (and $(d,p)$ respectively) made in \cite{PRC77} the statement on practical impossibility to determine systematical errors for  obtained by them values of enumerated parameters. One can assume that this statement concerns not only Axel-Brink hypothesis \cite{Axel,Brink}, but also rather ordinary systematical errors in determination of absolute intensities of the total gamma-spectra. Id est, reliability of all the data obtained by them is not completely determined. Consequently, all conclusions made by authors \cite{NIM} on parameters of cascade gamma-decay can be wholly mistaken due to rather significant coefficients $K_{tr}>1$ of transport of experimental errors $\delta S$ of measured of the total gamma-spectra intensity onto  the required  functions $\rho$ and $k$. Increase in $K_{tr}$ is strongly caused by the fact that any ``first generation spectra" in the region of low gamma-quanta energies are small difference of two large values (moreover, one of them cannot be determined experimentally even in principle). 

The errors $\delta$ of function $S$ and its parameters $X=\rho$, $X=k$ in vicinities of their most probable values are connected by approximate matrix equation:

\begin{equation}
\delta S=J \delta X J^{tr},
\end{equation}
where  $J$ - the matrix of derivatives of nonlinear function $S$ on its parameters $\rho$ and $k$. Matrix equation (1) can be solved on any modern computer. The program for calculation of Jacobi matrix $J$ (of derivatives $dS/dX$) in analytic form for the system of nonlinear equations solved in \cite{NIM} was prepared and tested in \cite{Gauss-M}. The sole problem by work with this program is caused by the only circumstance -- level density in diapason of neutron binding energy changes by 4-5 orders of magnitude. That is why, specific of computer arithmetic noticeably influences on results of calculation.

The presence of very significant non-diagonal elements in characteristic matrix 
$L=J^{tr} W J$ of the likelihood function for the system of equations solved in  \cite{NIM}, causes their degeneration for any possible weight matrix $W$ even by the use of additional data on density of neutron resonances, low-lying levels and by fixation of values of the total radiative widths of neutron resonances. Therefore, the desired $\rho$ and $k$ parameters  and their errors have a multitude of values with equal probability. This circumstance was not pointed out in the all papers performed with the use of method \cite{NIM}.

Authors of  \cite{PRC77} postulated that the error in level density derived from the spectra of evaporated nucleons does not depend on error in level density obtained with the help of \cite{NIM}. Id est., they did not take into account a possibility of strong correlation between systematical errors in determination of $\rho$. This correlation is stipulated, for example, by the use of identical hypothesis Axel-Brink \cite{Axel,Brink} for prediction of gamma-quanta emission partial widths and Bohr-Mottelson  \cite{BM} -- for nucleon products of nuclear reactions.

\section{Estimates of increase in errors of $\rho$ and $k$ at them extraction
from the total gamma-spectra}\hspace*{16pt}

Consequently, a multitude of equal-in-probability vectors $\delta X$ must be found from solution of equation (1) for matrixes $J$ corresponding to different equal-in-probability vectors $X$. The width of interval of the possible values of the vector elements $\delta X$ can be considered as the measure of experimental errors of $\rho$ and $k$.

The necessary for this operation vector $\delta S$ (or its upper estimates) in any experiment is determined only by analysis of factors carrying systematical distortions in the measured spectra. Its correct estimation can be done only by the authors of ``method Oslo" \cite{NIM} by means of corresponding calculations and additional experiments. For the present, this circumstance does not allow one to use equation (1) for estimation of region of the possible $\delta X$ vector values. Therefore, qualitative estimation of the expected systematical errors of $\delta\rho$ and $\delta k$ can be made only in the other way -- by comparison of the difference of the total gamma-spectra intensities at decay of levels from any spin window $J_1 <J<J_2$ for arbitrary excitation energy $E_{\rm ex}$, with difference of functional dependencies of $\rho$ and $k$ for pairs of their different model or experimental data sets used in the calculation. 

Naturally, this brings to loss of dependence on all the distortions carried in \cite{NIM}  at intermediate steps. The last concerns not only the procedure for determination of ``the first generation spectra", but and the use of mistaken (this is pointed out in first time in \cite{PEPAN-2005}) Axel-Brink hypothesis \cite{Axel,Brink} and, to a very high extent, to the method used for normalization of the total spectra for arbitrary energy $E_{\rm ex}$ to the same number of decays. Therefore, all the conclusions obtained below give notions only of the biggest permissible value (by using \cite{NIM}) of systematical errors in $S$.

The considerations presented above permit one to formulate main conditions and criterions for solution of the problem of partial evaluation of the error transfer coefficients of the measured total gamma-spectra onto the values of level density and radiative strength functions derived from them. First of all, there must be whose parameters  $\rho$ and $k$ were determined by two independent and really in principle different methods in the excitation energy interval being maximally close to neutron binding energy. It is also desirable to have the $\rho$ and $k$ values in one of analyzed sets with minimum possible uncertainty estimated by independent and traditional methods of error determination. They must be extracted from experiment without the use of untested hypotheses (like \cite{Axel,Brink,BM}). It is also desirable to have the maximum different types of experimental data for functional dependencies $\rho$ and $k$ on excitation and gamma-quantum energies. The mentioned requirements are satisfied, for example, in compound nuclei $^{96}$Mo and $^{172}$Yb. The total gamma-spectrum at maximum excitation energy $E_{\rm ex}$ is (to a precision of different population of initial levels with different $J^{\pi}$) a superposition of the ``first generation spectra" for this and lower excitation energies.
That is why, one can expect that the estimated below effect of influence of the total gamma-spectra systematic errors on uncertainties of desired  $\rho$ and $k$ values has no principle differences relative the case \cite{NIM}.

\section{Conditions of calculation}\hspace*{16pt}

The total gamma-spectrum following decay of levels from narrow interval
$\delta E_{\rm ex}<<E_{\rm ex}$ for any excitation energy bin of a nucleus under study can be normalized to both the given decay number of initial levels or to the total cascade energy $E_{\rm ex}$. In the first case, systematical error $\delta S/S$ is an algebraic sum of errors in determination of form of measured spectra at different energies of gamma-quanta and inevitable error of absolute normalization. In the second case:

\begin{equation}
\sum I_\gamma E_\gamma=E_{\rm ex}                                    
\end{equation}
all the systematical errors are minimal and have sign-changeable dependence on $E_\gamma$ and in sum are equal to error $\delta E_{\rm ex}$. The use in calculation of different functional dependences for $\rho$ and $k$ brings (at the of normalization (2)) to analogous effect. Its value can be characterized by the parameter 
$D^2=\sum(2(S_E-S_T)/(S_E+S_T))^2$ for any pairs of the ``standard" and ``tested" total gamma-spectra $S_E$ and $S_T$, respectively.

It is postulated below that any sets of $\rho$ and $k$ can be obtained with equal probability from analysis like that in \cite{NIM}, if only mean-square difference $D$ of the total gamma-spectra calculated with the use of them is approximately the same and equals square root mean of errors of the experimental spectrum. Correspondingly, reliable determination of experimental values of desirable parameters  $\rho$ and $k$ requires that mean-square error of experimental spectra must be much less than mean-square difference $D$ of two corresponding spectra.  
The existing differences in the $\rho$ and $k$ values give notion on real magnitudes of their systematical errors. In practice, extraction of the $\rho$ and $k$ values from $S$ is performed in several steps \cite{NIM}. Therefore, each step brings to additional increase in systematical errors of the determined values. This circumstance must be taken into account at estimation of required precision at determination of $S$.

Comparison between the data on $\rho$ and $k$ obtained in Oslo within the method \cite{NIM} and Dubna data  \cite{PEPAN-2005} allows us to reveal their characteristic peculiarities and to determine the minimal
permissible 
systematical error $\delta S$ which guaranties reliable identification of minimal difference of the tested parameters. It is determined, first all, by the presence of the step-like structure and correlated with it in position peak in the radiative strength functions in the Dubna data. The Oslo data on level densities are very close to the existing primitive models (for example, \cite{BSFG}). Unlike the Dubna data they do not show abrupt change in nuclear properties below neutron binding energy.

Practically, there were used in calculations the models of radiative strength functions\cite{Axel,KMF}  for $E1$-transitions, \cite{BSFG,Strut} -- for level density, results of approximation of Dubna parameters of cascade gamma-decay \cite{PEPAN-2006,Appr-k} and the experimental data on $\rho$ and $k$ for $^{96}$Mo $^{172}$Yb,  obtained in Oslo \cite{Mo96,Yb172} for reactions induced by $^3$He. Comparison between results of calculation for combinations:
\\
1. \cite{KMF,BSFG},\\
2. \cite{Axel,BSFG},\\
3. \cite{KMF,PEPAN-2006},\\
4. \cite{Axel,PEPAN-2006},\\
5. \cite{KMF,PEPAN-2006,Appr-k},\\
6. \cite{Axel,PEPAN-2006,Appr-k},\\
7. \cite{Mo96,Yb172}\\

was performed for the energy of initial level $E_{\rm ex}=B_n$ for two its possible parities and spins excited at  the thermal neutron capture. Below 
$E_d=2.5$ MeV for $^{96}$Mo and 1.85 MeV for $^{172}$Yb in calculation was used experimental information \cite{ENSDF} on decay modes and parameters of known levels of these nuclei. Parameters of the model \cite{Strut} were chosen from those obtained in \cite{PEPAN-2006} in such a way that to have smallest discrepancy with \cite{BSFG}. Information on energy resolution of scintillation detectors was included in calculation as well.

It was supposed in all calculations that the level densities of different parity above the energy $E_d$ are equal; only dipole transitions were taken into account. The sum of $k(E1)+k(M1)$ in calculation of parameter $D_k$ for different strength functions was re-normalized so that the total radiative width of decaying level was equal in all variants of calculation. All the total gamma-spectra were normalized to the energy $B_n$ (in correspondence with (2)). The total area of calculated spectrum at normalization (2) to the cascade energy is obviously equal for any set of $\rho$ and $k$.

Therefore, all the calculated variants of spectra differ each from other only by shape of dependence $S=f(E_\gamma)$. Just this circumstance decreases sensibility of the method like \cite{NIM} as compared with extraction of $\rho$ and $k$ from the two-step cascade intensities. This conclusion is true only for the data for $\rho$ and $k$, obtained in Dubna. The results of the cascade intensity analysis performed in Prague (and by other groups -- in the same manner) are not reliable due to the presence in corresponding method of three principle mistakes. This was shown in \cite{Prep134}.

It is naturally that the forms of calculated spectra depend also on ratio between excitation probability of initial levels with different $J^{\pi}$, on difference in level densities with different parity, ratio between strength functions of dipole gamma-transitions of different type and so on. But, judging by the data listed below, limitation in extent of variation of calculation parameters by the mentioned upper variants cannot bring to radical change of the made below conclusions on required precision in determination of $S$. 

At calculation of concrete values of parameters $D$ there were introduced the low and high thresholds for $E_\gamma$. They were equal to 0.5 and 7.0 MeV for molybdenum  and 0.5 and 6.0 MeV -- for ytterbium. This was done for reduction of contribution in value $D$ of parts of spectra with low intensity.

Seven obtained variants of the calculated total gamma-spectra for these isotopes are shown in Fig. 1. Their parts corresponding to spectra of only primary gamma-transitions are presented in Fig. 2 in the same scale. It is natural that the ratio between intensities of the spectra presented in figures 1 and 2 practically does not depend on the used method of normalization. The used in the calculation level densities (excited by the primary dipole gamma-transitions following thermal neutron capture in nuclei under study) are shown in Fig. 3, and strength functions -- in Fig. 4. The ratios $r$ intensity of the calculated total gamma-spectra to intensity of only their primary transitions is presented in Fig. 5.

\section{The Axel-Brink hypothesis testing problems}\hspace*{16pt}

Modern nuclear models take into account influence of structure of levels connected by gamma-transition on its probability. That is why a necessity to use the Axel-Brink hypothesis \cite{Axel,Brink} is absent in modern theory. But, a precision of any theoretical calculation is determined by obviously insufficient accuracy of the experimental data which were the basis for development and parameterization of the model like QPNM \cite{QPNM}. Unambiguous conclusion on existence of strong dependence $k=\phi(E_\gamma, E_{\rm ex})$ was firstly obtained from the experiment \cite{PEPAN-2005}. Unfortunately, due to methodical reasons information on this function is limited by the region $E_{\rm ex} \approx 0.5B_n$. The degree of mistakenness  of the Axel-Brink hypothesis at high excitation energy can be estimated only by experimental study of cascades with three and more successively emitted gamma-quanta. Corresponding experiment was realized by authors of \cite{Fur07}.

In spite of limited character of the data \cite{PEPAN-2005}, they allow one to expect a possibility of very strong or even complete compensation of decrease of level density at any excitation energy $E_{\rm f}$ with respect to general trend by increasing of the gamma-transition widths directly follows to corresponding levels. It was shown in \cite{BM-err} that the analogous effect should be taken into account and in analysis in the spectra of evaporated nucleons also.

There are no principle difficulties to calculate the total gamma-spectrum under assumption of complete compensation of decrease in level density by simultaneous increase in strength functions. In the limit of the possible complete 
$\delta \rho/\rho=-\delta k/k$  compensation of deviations of parameters of gamma-decay from general trend the calculation of the total gamma-spectrum with or without accounting for hypothesis \cite{Axel,Brink} must give the same total gamma-spectrum. In the other words, the method \cite{NIM} is weakly sensitive or completely insensitive to shape of function $k=\phi(E_\gamma, E_{\rm ex})$.      

But, cascade population of any level at any excitation energy depends on the cross-section of gamma-quantum interaction with excited nucleus and is not compensated by change in level density. Therefore, analysis of cascade intensities  \cite{PEPAN-2005} allows one to study real dependence of strength functions on both gamma-quantum energy and structure of nucleus. The method like \cite{NIM,Bar} has not such possibilities.

\section{Results of comparison of calculation}\hspace*{16pt}

The results of comparison of mean-square relative differences of the functions under consideration are listed in Table. The parameters $D$ for all combinations were determined at choice of the $\rho$ and $k$ values as the ``base" first variant. This combination is more often used in different calculations.

{\sl Table. The square mean differences   of intensities of calculated total gamma-spectra $D_s$, intensities of only primary gamma-transitions $D_{pr}$, the used level densities $D_\rho$ and radiative strength functions $D_k$. $M_s$, $M_{pr}$,  $M_\rho$ and $M_k$} -- the modules of their maximal relative differences.

\begin{center}
\begin{tabular}{|l|r|r|r|r|r|c|r}  \hline
Parameter&\multicolumn{6}{|c|}{Variants}\\ \hline
$^{96}$Mo & 1,2 & 1,3 & 1,4  & 1,6  &  1,5 &   1,7\\ \hline
$D_s$   & 0.09 & 0.28 & 0.28 & 0.24 & 0.24 &  0.24\\
$M_s  $ & 0.43 & 0.80 & 0.80 & 0.59 & 0.56 &  1.13\\
$D_{pr}$& 0.18 & 0.35 & 0.30 & 0.31 & 0.24 &  0.40\\
$M_{pr}$& 0.58 & 0.81 & 0.75 & 0.74 & 0.61 &   1.6\\
$D_\rho$&  0   & 0.56 & 0.56 & 0.56 & 0.56 &  0.56$^{*}$\\
$M_\rho$&  0   & 1.3  & 1.3  & 1.3  & 1.3  &  1.3 \\
$D_k$   & 0.21 & 0.  & 0.20 & 1.04  & 1.07 & 0.49 \\
$M_k$   &  0.7 & 0.  & 0.70 & 1.62  & 1.63 & 1.46\\
\hline
$^{172}$Yb& 1,2& 1,3  & 1,4  & 1,6  &  1,5 & 1,7 \\\hline
$D_s$   & 0.12 & 0.19 & 0.12 & 0.25 & 0.22 & 0.22\\
$M_s$   & 0.49 & 0.68 & 0.35 & 0.80 & 0.57 & 0.63\\
$D_{pr}$& 0.16 & 0.21 & 0.27 & 0.22 & 0.21 & 0.26\\
$M_{pr}$& 0.59 & 0.64 & 0.75 & 0.45 & 0.44 & 0.67\\
$D_\rho$& 0    & 0.49 & 0.49 & 0.49 & 0.49 & 0.03\\
$M_\rho$& 0    & 1.09 & 1.09 & 1.09 & 1.09 & 0.04\\
$D_k$   & 0.17 & 0.0  & 0.17 & 0.78 & 0.78 & 1.28\\
$M_k$   & 0.64 & 0.0  & 0.64 & 1.23 & 1.24 & 1.67\\
\hline
\end{tabular} \end{center}
{\sl $^{*}$Experimental level density of $^{96}$Mo, given in \cite{Mo96}, completely corresponds to model \cite{BSFG}.}

These data allow one to get some notion on extent of increase in spectrum error of the primary gamma-transitions used in \cite{NIM} for determination of the $\rho$ and $k$ values. To a precision of coefficient $\sim 1.4$ (or some more), the data in Fig. 5 gives an extent of increase of error of total gamma-spectrum intensity at its transformation in the spectrum of only primary gamma-transitions. Coefficients of increase in systematical errors of $\rho$ and $k$ owing to the use of Axel-Brink hypothesis instead of the radiative strength function depending on energy of excited level (structure of its wave function) cannot be less than 2 (obtained from comparison of $\rho$ for the same nuclei in \cite{Meth1} and \cite{PEPAN-2005}). It can have arbitrary  value. The way for its observation -- variation of forms of the initial $\rho$ and $k$ functions as in the method \cite{NIM} and using only library programs of multidimensional fitting.

\section{Analysis of results}\hspace*{16pt}

Maximal value of $D_s$ for the parameters used by the testing does not exceed 0.25 -- 0.28. Correspondingly, maximal error in measurement of the total spectra must be 3-5 times less. The procedure of determination of the ``fist generation spectra" brings to additional increase of experimental errors in the region above threshold of the experiment in 3-10 times and more (Fig. 5).

In practice, infringement
of the Axel-Brink hypothesis appears itself in simultaneous increase in $k$ at decrease in $\rho$ values in the region of the step-like structure. This effect was revealed for both primary and secondary gamma-transitions. But, observation of compensation of decrease in $\rho$ by increase in $k$, as it was obtained in \cite{PEPAN-2005}, requires of both rise of precision at determination of intensity of total spectrum and to decrease of the $D_s$ value. Corresponding coefficient in our estimation can be adopted to equal to that presented above. As a result, one can assume in the first approach that certain and unambiguous identification of different values of level densities and strength functions requires that the total relative error of determination of absolute intensity of the total gamma-spectra in method \cite{NIM} must be, most probably, less than 0.01 at least for low energy bins of the measured spectra.

This conclusion quite unambiguously from the made in \cite{TSC-57,BecK08,TSC-172} attempts to reproduce intensity of two-step cascades in $^{57}$Fe, $^{96}$Mo and $^{172}$Yb by means of level density and strength functions derived by ``Oslo method".
There are made in publications the attempts to reproduce the sum of unknown two-step cascade intensities with close values of energies of the primary and secondary gamma-transitions by calculations with corresponding pairs of $\rho$ and $k$. Impossibility to prove unambiguous reproduction of each of these items is obvious even if it was achieved for the sum and for all energies of the cascade gamma-transitions. In the other case -- the data \cite{NIM} moreover cannot correspond to the $\rho$ and $k$ values which reproduce intensity of cascades.

Analysis of these data \cite{Prep134,Fe57,Oslo-err} performed in correspondence with requirements of mathematical statistics and mathematics completely excludes reliability of determination  of $\rho$ and $k$ by method \cite{NIM}, at least, for the gamma-decay of compound states excited at thermal neutron capture. Moreover, this question was stated in details in \cite         {Prep134,Fe57,Oslo-err}.

\section{Conclusion}\hspace*{16pt}
So, the method \cite{NIM} for determination of the $\rho$ and $k$ values can give reliable information on nuclear parameters only in the experiment which provides rather high precision of measurement of the total gamma-spectra. It is probable that the spectrometer used for this aim in Oslo cannot provide for required precision even in principle.

\begin{figure}
\leavevmode\hspace{-.8cm}
\epsfxsize=15cm

\epsfbox{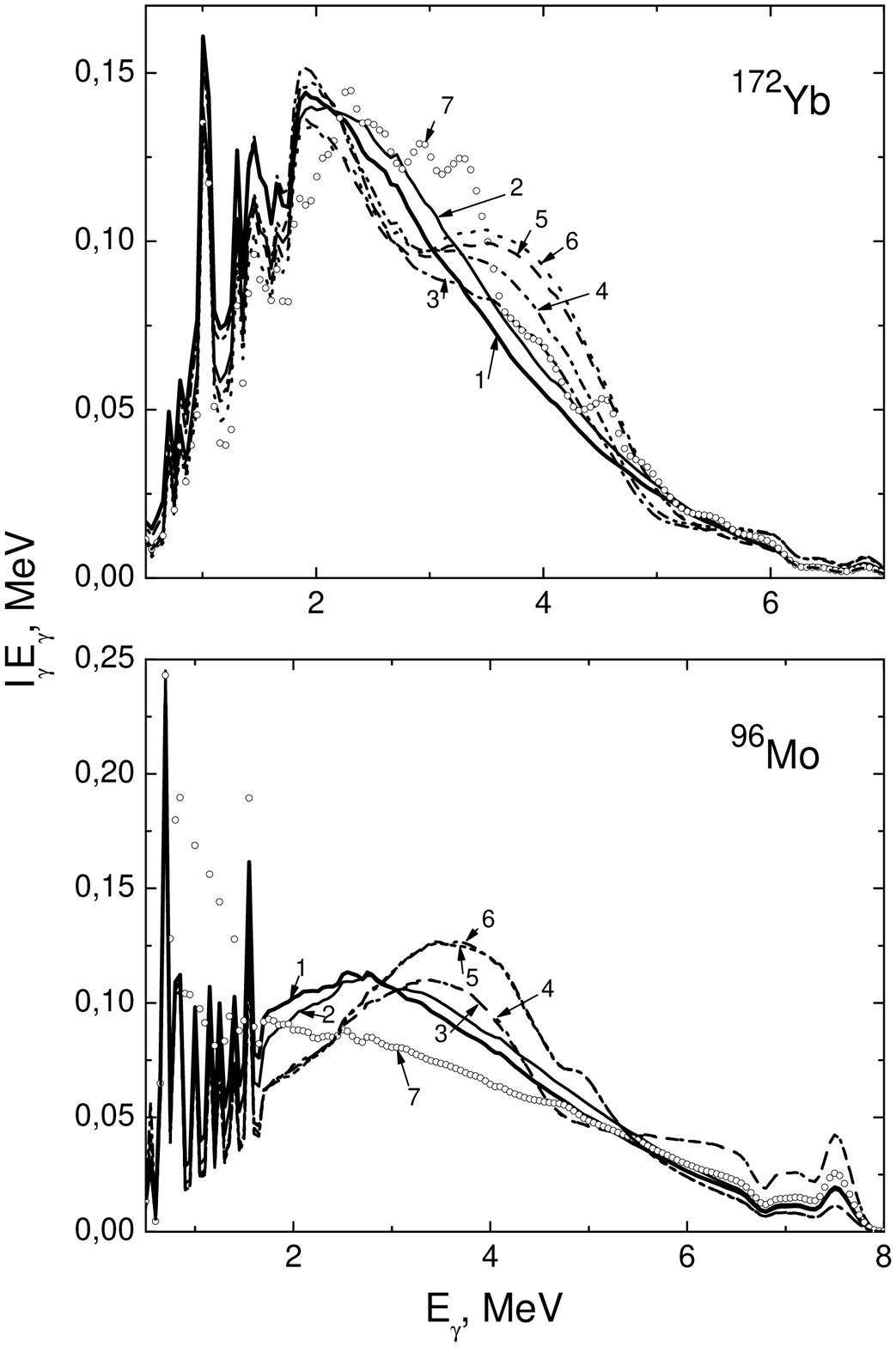}

{\bf Fig.~1.} The calculated total gamma-spectra of $^{96}$Mo and $^{172}$Yb
for seven combinations of the data on $\rho$ and $k$. Figures -- number of calculation variant.

\end{figure}
\begin{figure}
\leavevmode\hspace{-.8cm}
\epsfxsize=15cm

\epsfbox{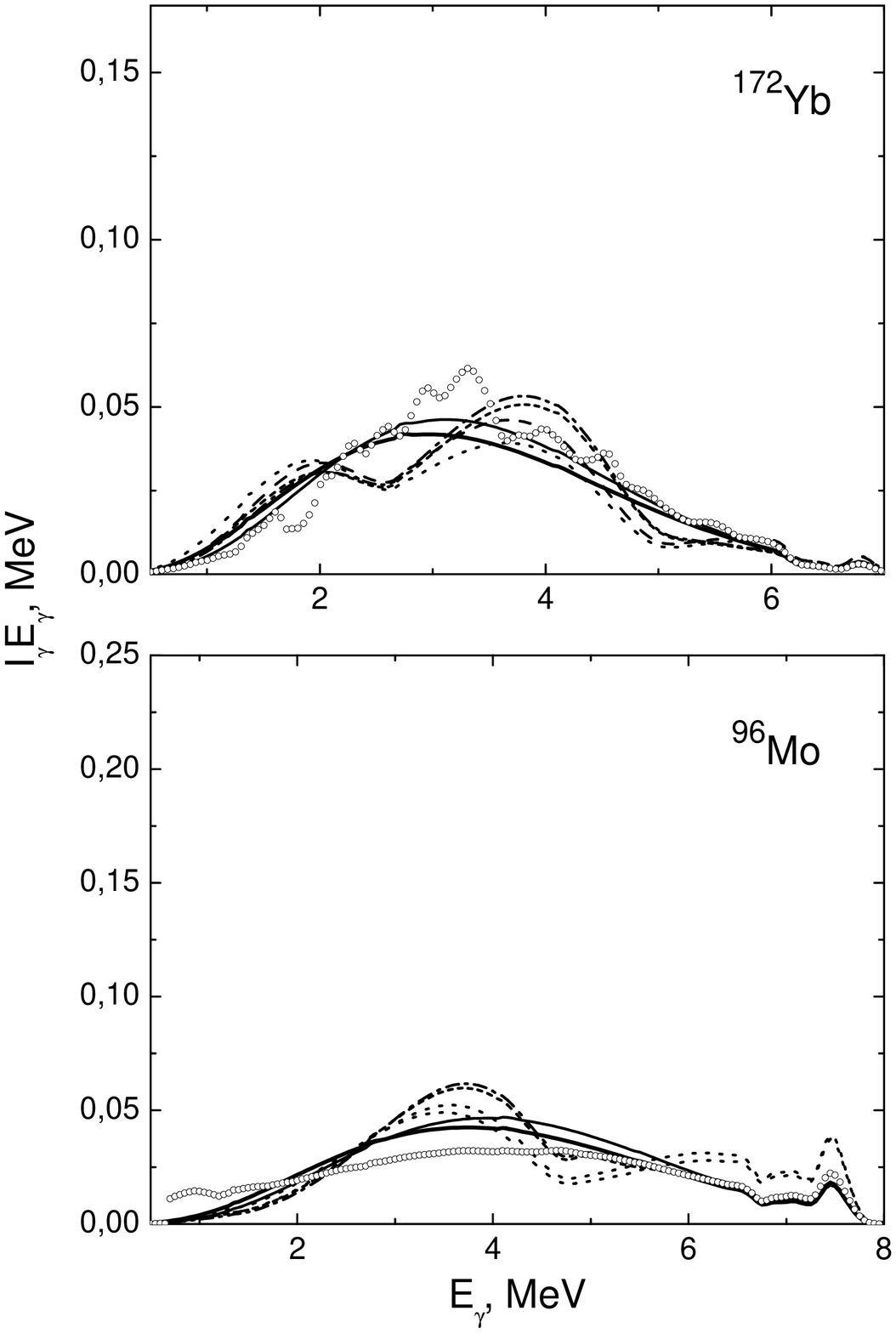} 

{\bf Fig.~2.} The same, as in Fig.~1, for corresponding spectra of the primary gamma-transitions.
\end{figure}
\begin{figure}
\leavevmode\hspace{-.8cm}
\epsfxsize=15cm
\epsfbox{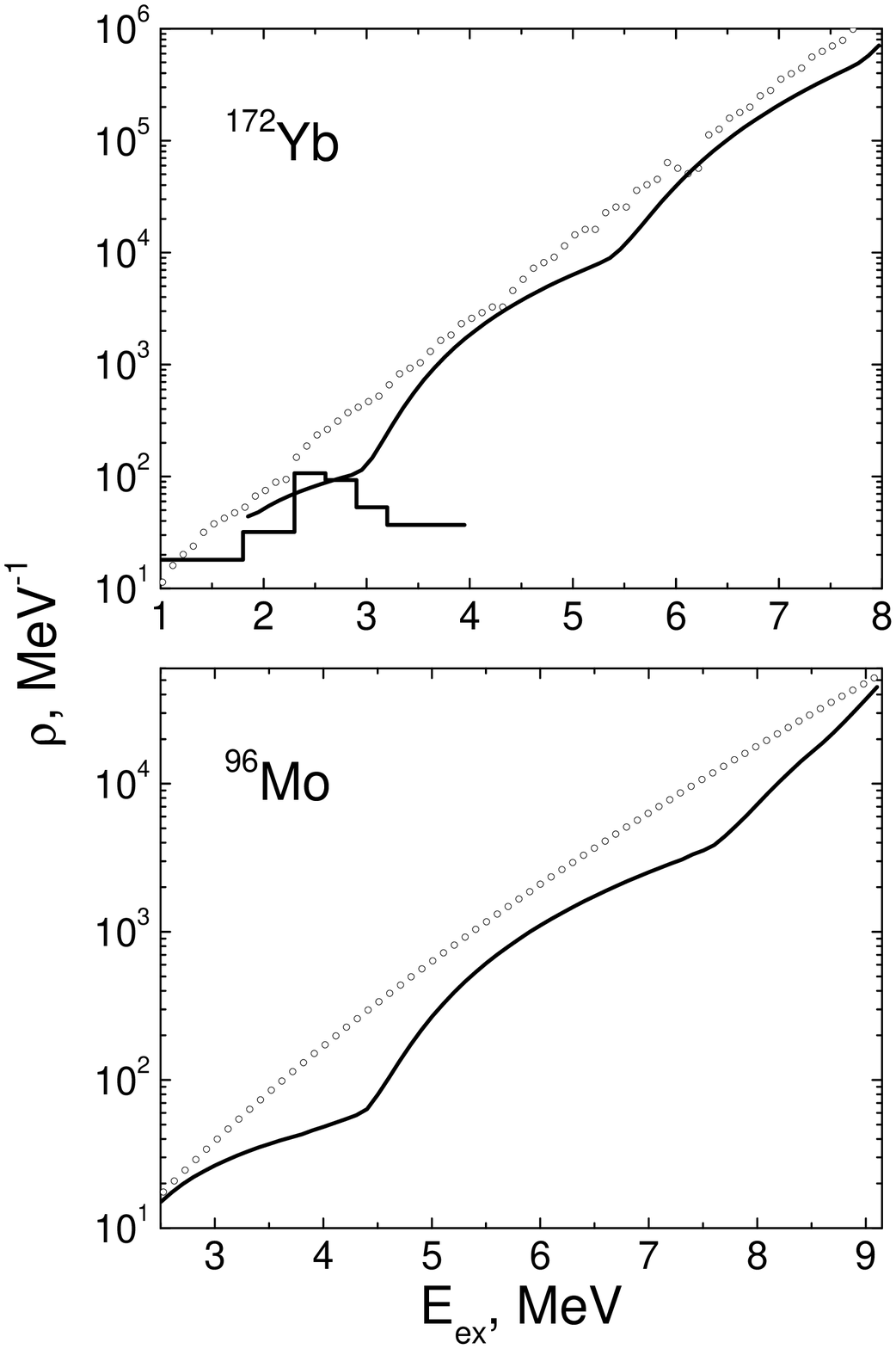}

{\bf Fig.~3.} The accepted in calculation level density excited by the dipole primary gamma-transitions at the thermal neutron capture. Points -- data \cite{Mo96,Yb172}, lines -  \cite{Strut}, histogram -- level density from reaction $(\overline{n},\gamma$), obtained by analogy with \cite{Avres}.
\end{figure}

\begin{figure}
\leavevmode\hspace{-.8cm}
\epsfxsize=15cm

\epsfbox{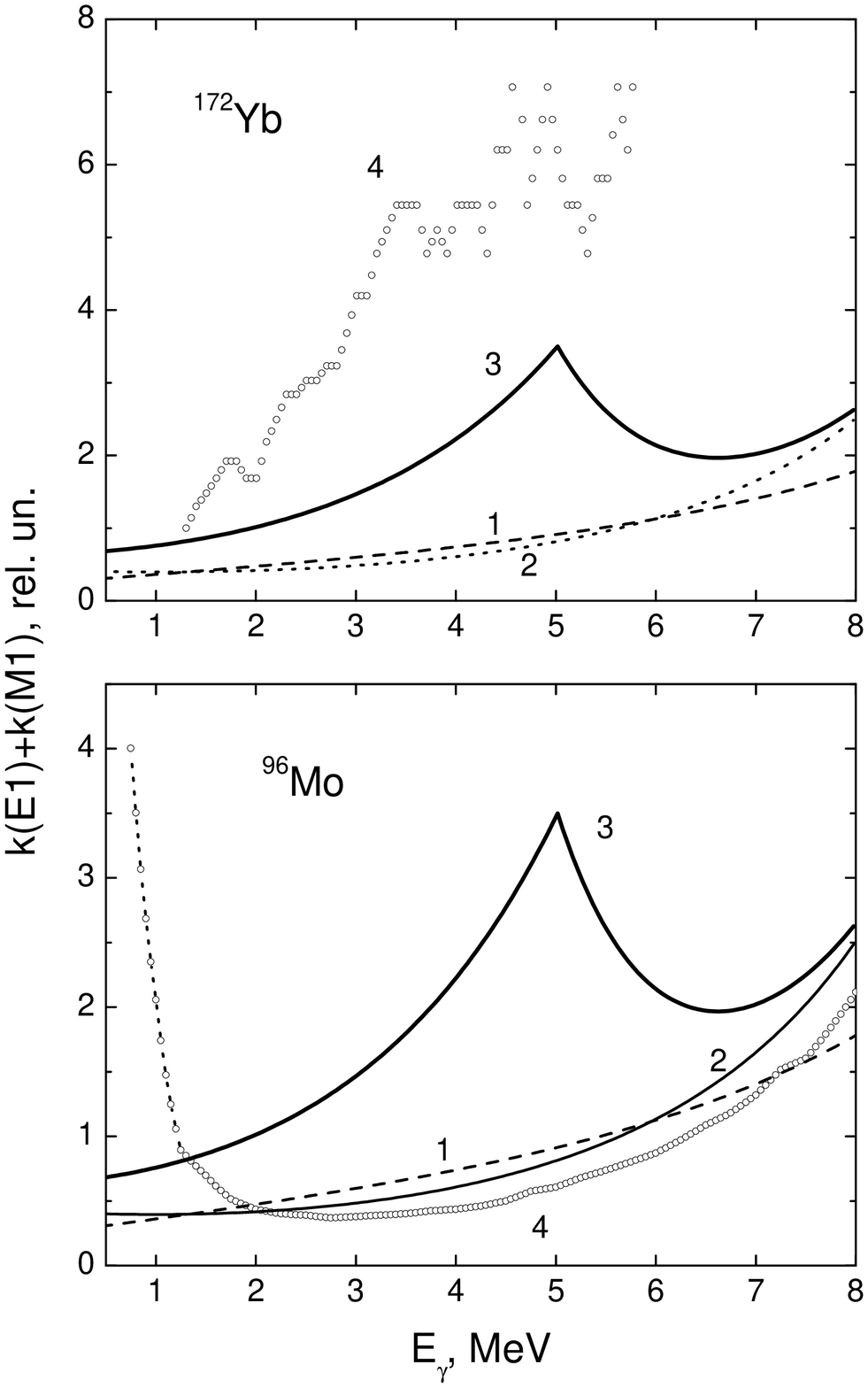} 
{\bf Fig.~4.} The same, as in Fig.~3, for the used $k$ values. Lines: 1 -- \cite{Axel}, 2 -- \cite{KMF}, 3 -- variant with the use of approximation function from  \cite{Appr-k}. Points -- data \cite{Mo96,Yb172}.
\end{figure}
 \begin{figure}
\leavevmode\hspace{-.8cm}
\epsfxsize=15cm

\epsfbox{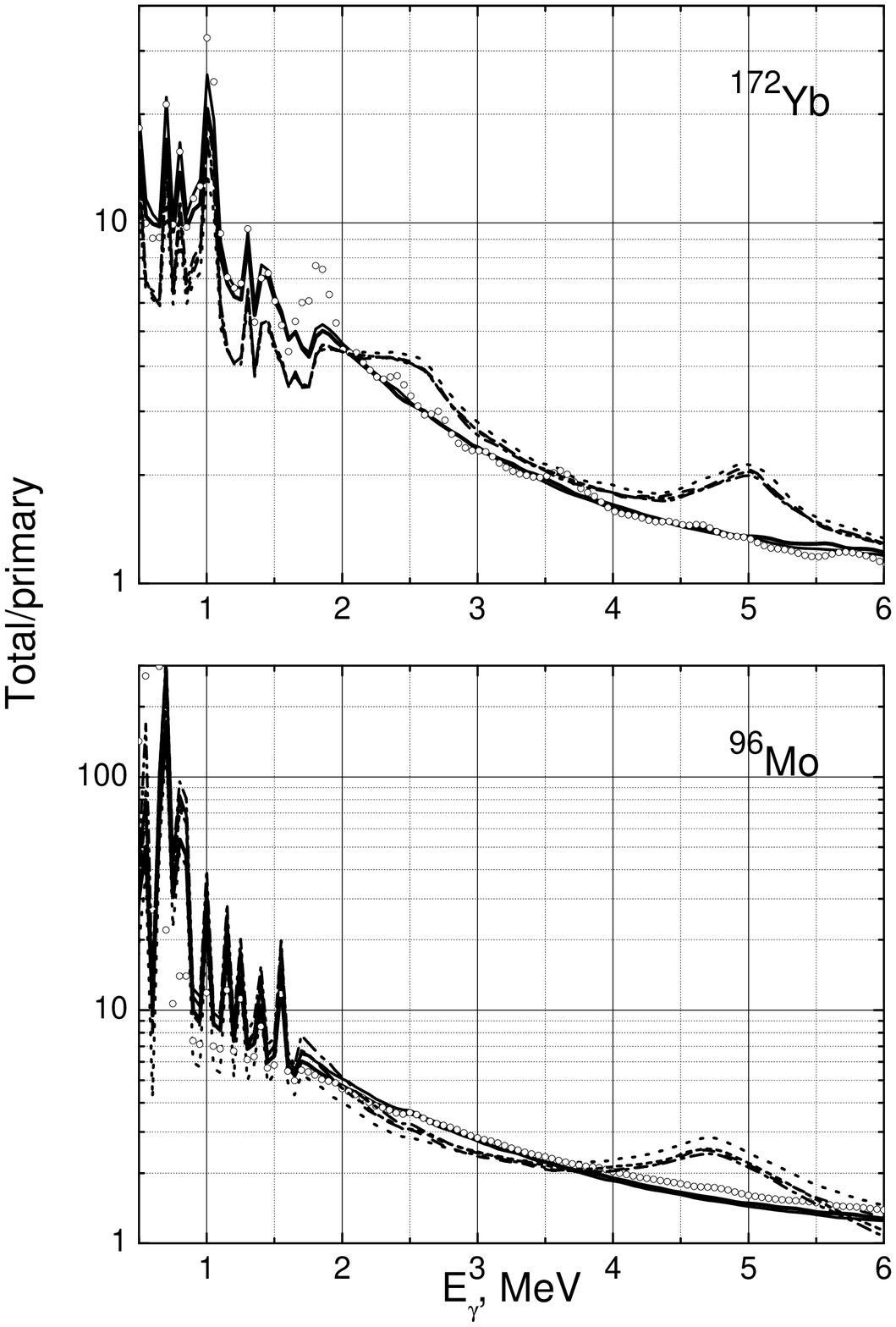}

{\bf Fig.~5.} The ratio of the total gamma-spectrum intensity to intensity of only primary gamma-transitions.
\end{figure}

\end{document}